# Computer Science Communities: Who is Speaking, and Who is Listening to the Women?

## Using an Ethics of Care to Promote Diverse Voices

Marc Cheong, Kobi Leins, Simon Coghlan
Centre for AI and Digital Ethics (CAIDE), Faculty of Engineering and IT (FEIT)
University of Melbourne
Parkville VIC Australia
{marc.cheong, kleins, simon.coghlan} @ unimelb.edu.au

## ABSTRACT

Those working on policy, digital ethics and governance often refer to issues in 'computer science', that includes, but is not limited to, common subfields of Artificial Intelligence (AI), Computer Science (CS) Computer Security (InfoSec), Computer Vision (CV), Human Computer Interaction (HCI), Information Systems, (IS), Machine Learning (ML), Natural Language Processing (NLP) and Systems Architecture. Within this framework, this paper is a preliminary exploration of two hypotheses, namely 1) Each community has differing inclusion of minoritised groups (using women as our test case); and 2) Even where women exist in a community, they are not published representatively. Using data from 20,000 research records, totalling 503,318 names, preliminary data supported our hypothesis. We argue that ACM has an ethical duty of care to its community to increase these ratios, and to hold individual computing communities to account in order to do so, by providing incentives and a regular reporting system, in order to uphold its own Code.

## KEYWORDS

Gender, Diversity, Computer Science, publishing, research, sex equality, gender representation.

## 1 Introduction

Lack of diversity in research and development of computing technologies has long been a well-recognised problem [10,16,47]. This persisting problem concerns both questions of fairness and the quality and breadth of research. Homogeneity in computer science communities is the most the recent target of a slew of research. [15, 52] Some argue that decolonisation of computing and big data – which is amplified at speeds and scales previously unimaginable – needs to be given much greater regard within computational communities [5,33]. Although the sciences are often thought to be neutral and unbiased, the lack of diversity and of a variety of different voices in scientific fields is increasingly being researched — not just for the sake of individuals but also in the interests of maximising the value and quality of research in those fields. The term *minoritised*, rather than *minority*, is used because those represented are not necessarily in the minority more widely speaking, but rather are effectively kept in the minority for a wide range of reasons, usually by a dominant group [12]. To explore these issues, we turn to the 'ethics of care', in part because of its emphasis on closely attending to the experiences and needs of the minoritized. Using this care ethics approach, we argue that there is an duty of care and justice to increase diversity within computer science communities, and that those with the power need to act to include these people and voices in computational sciences. In this paper, we focus on women in computer science. However, we also suggest that our broad findings about duties of care and justice can be applied to other minoritized groups. We also argue that, in addition to meeting ethical responsibilities, creating more equitable representation can strengthen the disciplines themselves.

Lack of diversity can affect both natural sciences and the humanities. For example, a study on research into birdsong unearthed data that showed that research had been skewed towards male birdsong until female researchers focuses also on female birds [19]. Similarly, in philosophy, "feminist methods of articulating ethical theories" in areas such as moral philosophy have been few and far between until the emergence of feminist ethics in the 1970s and 1980s [40]. In tertiary education, female academics are often judged much more harshly by male and female students alike. [6,30,31,34,43]. Although this question is relevant for a much broader group of non-mainstream computer scientists, gender is the easiest (within a margin of error) to identify and map empirically. For this reason, we have started the discourse in an area where we could perform automated quantitative analysis. This paves the way for future research involving, for example, interviewing experts within each of the sub-fields to gain a clearer understanding of their cultures and drivers.

In computer science, the statistics [1] are not encouraging: the "number of women studying computer science is falling… since the 1970s" [35], with numbers "… falling pretty steadily since the 80s, despite the increase in demand for these types of skills" [35]. In the broader picture, "women in STEM [Science, Technology, Engineering, and Mathematics] make $16,000 less on average than their male counterparts" [35]. To date, the only breakthroughs include one recent paper has been written about the role that gender plays in senior roles and publishing within HCI [32].

However, there is a gap in extant research on the status of female academics *across* computer science communities which look at measures of 'impact' in the context of citations of ranked conferences and papers. Although there are many ways to assess diversity, such as comparing the number of employees in a field or the number of graduates at a University, we choose to take the approach of using publication rates. Given the number of fields we are analysing, and the volume of the data available, we chose this last approach as an entry point to start investigating the state of the different computer science fields relative to each other.

We wish to stress upfront that gender identification is problematic and may cause representational harm, and that sexuality is non-binary, an issue surveyed in some detail by Saif M Mohammad [37].

**There are broad methodological caveats that we would like to address upfront in the spirit of transparency:**

1. Extant algorithms assume gender is a binary construct, and hence do not account for diverse gender identities. Reiterating a point in [37]: "Gender is complex, and does not necessarily fall into binary male or female categories (e.g. nonbinary people), and also does not necessarily correspond to one's assigned gender at birth" [37].

2. Most systems for gender determination are deterministic based on prior statistics of name usage. Hence, there is no other context [36] – e.g. an individual's confirmation of preferred gender marker – beyond the mere isolated first name. Such binary gender research omits many minoritised groups. A "strong normative tendency to use names to signal gender" can also lead to misgendering – "a machine associat[ing] someone with a gender with which they do not identify" – which causes harm [37][2].

3. Statistics on first names are biased towards popular/frequently-occurring names in a mostly anglophone context, and are limited to the cultural and temporal context when the list is produced. (e.g. the USA Social Security Administration Baby Names dataset).

We note that although we are focusing on women, our concerns include broader intersectionality and representation within the field. We are also concerned about marginalisation of all those whose voices need to be amplified in the various fields of computer science. We have chosen to use the blunt instrument of automated identification of gender based on names not because it is without risk, but because it is critical that we ensure diversity to ensure more fair, accountable and transparent systems, and this is one method to provoke that debate. This is not a study of individuals, but an attempt to provoke discussion on the basis of a broad stroke analysis that, although problematic, we suggest is useful to move the conversation, and ethics of care and justice, forward to ensure inclusive and better research in computer science. Hence, in this paper, we seek to answer these four initial research questions.

**RQ1.** How are female academics currently represented in publishing in each of the subfields of computer science?

**RQ2.** How do the statistics of publications compare to female representation in each computing science community, based on existing research? What differences, if any, are there between different computer science communities?

**RQ3.** What might ethics, and specifically an ethics of care, have to say about underrepresentation of women in computer sciences?

**RQ4.** How can we reduce the disparity?

We organise this paper as follows. First, the literature of gender bias in computer science and associated fields are surveyed to ascertain the current state of gender diversity. Second, we detail our experimental methodology: the choice of subfields surveyed in our experiment; our data sources and data analysis choices; and ethical considerations. Third, we analyse our results based on extant surveys of subfields, and those found in our initial quantitative and qualitative research. Fourth, we introduce an "ethics of care"

---

[1] Besides having a gender disparity, computer science still has a disparity in enrolments based on ethnic background: an average of $14,000 [35] difference between White Americans and Black/Hispanic Americans in STEM fields.

[2] See also Catherine Connell. 2010. Doing, Undoing, or Redoing Gender?: Learning from the Workplace Experiences of Trans people. *Gend. Soc.* 24, 1 (Feb. 2010), 31–55;

also:
Foad Hamidi, Morgan Klaus Scheuerman, and Stacy M Branham. 2018. Gender Recognition or Gender Reductionism? The Social Implications of Embedded Gender Recognition Systems. *In Proceedings of the 2018 CHI Conference on Human Factors in Computing Systems* (Montreal QC, Canada)(CHI '18,Paper 8). Association for Computing Machinery, New York, NY, USA, 1–13 framework, and explore some of its implications in computing sciences. Finally, we introduce some recommendations and explain what further research is necessary.

## 2 Literature Review

Some research documents the gender disparity in the sciences including the computer sciences. A 2001 qualitative study at Stanford University found women only made up 9.75% of Computer Science professors [2]. Causes identified by the researchers include "motivation, parental support, balancing family and personal life, any perceived gender biases or discrimination against women, the enticement of industry versus academia, and the views of both women and men towards women in computer science" [2]. In a recent study on academic publishing statistics across "83 countries and 13 disciplines" [25], Huang et al investigated citation data of 1.5 million authors, with their gender identified from academic authorship records on Web of Science, and independently replicating the results against the Microsoft Academic and DBLP databases. Their findings indicate that "…paradoxically, the increase in the number of women academics over the past 60 years has increased … gender differences in the total productivity and impact of academic careers" [25] in STEM. This matches findings in other studies [35].

Extrapolating to future authorship trends in computer science, Wang et al [61] conducted an analysis on 2.87 million computer science academic publications since the 1970s, using name-based gender inference and time series forecasting. The forecast is rather grim: "based on recent [gender] trends, the proportion of female authors in Computer Science is forecast to not reach parity in this century" [61]. This result agrees with the work of Holman et al, whose analysis of "36 million authors from >100 countries publishing in >6000 journals" [24] reveals a gender gap which is "likely to persist for generations… [and] clearly require[s] additional interventions if parity is to be reached this century" [24].

It should be noted that other dimensions – in particular local cultural and political context – can have an *intersectional* [33] negative effect on diversity. A study by Thelwall et al in the context of STEM in India, found that overall there is a "…substantial overall male bias, [but the] broad research field choice is *less* influenced by gender" [56]. In other words, male bias still exists overall, but the *distribution* of bias across subfields are different. Local factors play a role in changing this distribution: examples include a higher coverage of "algorithms in Indian mathematics [studies]…" and "a tendency for males to research thing-oriented topics and for females to research helping people and some life science topics" [56]. A 2008 qualitative study by Lagesen conducted in a Malaysian computer science faculty revealed that more than half the Bachelor students in computer science/information technology are female, around 40% of postgraduates are female, and encouragingly "the majority of the faculty, as well as all heads of departments and the dean, were women" [28]. It is clear that computer science is not inherently gendered – women *can* and *should* succeed given the right circumstances, which may include responsible action by the profession.

Many of the above studies refer to a *'pipeline effect'* [2,28] – i.e. to the likelihood of leaving academia as one progresses through the academic 'pipeline' from undergraduate years to tenure (see also [49]). There is also literature on the motivations and experiences of university undergraduates in pursuing (or otherwise) computer science-related courses. Sax et al.'s review of American university students' survey responses in their freshmen year (sampled from 1976 to 2011) found a "persistent, sizeable underrepresentation of women across all years" [46] in computer science. At the same time, the number of women in STEM and other professions more broadly has risen [38].

## 3 Methodology

The motivation for this research was to analyse publication rates by gender in individual communities within computer science, from 1969 to 2020 (inclusive). The reason is that much of the research on governance and ethical considerations regards computer science as a single area of research, when, in fact, it is constituted by many individual (sometimes overlapping) communities.

In designing this paper's methodology, we consulted existing literature on experimental methods or heuristics used to approximate gender representation in academia. The most frequently used heuristic is inferring gender from author names in publication records, accessible via academic citation databases [24,25,56,61]. Commonly used databases include DBLP, Microsoft Academic, Scopus, and Web of Science. In terms of the actual algorithmic methods used for name-to-gender inference, Santamaría and Mihaljević [45] have conducted a thorough literature review and benchmark on five such services. In brief, there are both offline techniques using open data (such as *gender-guesser* [42] for Python which is based on curated data sources [45]), and online techniques (web services or APIs which are proprietary in nature, such as *genderize.io* and *GenderAPI*). Offline techniques use simple statistics based on frequency of names, with the advantage of transparency and simplicity [9] but with the disadvantage of not being frequently updated or representative of the global population (e.g. not being able to infer culturally-diverse names). Online techniques are the inverse: they have a higher accuracy and inclusion of diverse names, but with the disadvantage of being a commercial offering without much transparency about their inner workings. The use of name-to-gender algorithms have ethical caveats and limitations, documented per our Introduction in Section 1.

### 3.1 Determining Subfields of Study

For this paper, we have identified nine subfields of computer science, based both on our own experience in the discipline and the identification of common research focus areas in top universities. These subfields represent common research themes in academic institutions with a computer science department (which included similar-sounding departments such as computer systems and information technology). These nine subfields are: Artificial Intelligence (AI), Computer Science (CS), Computer Security (or cybersecurity or information security, abbreviated InfoSec), Computer Vision (CV), Human Computer Interaction (HCI), Information Systems (IS), Machine Learning (ML), Natural Language Processing (NLP), and Systems Architecture (SA). The capitalised proper nouns are used to uniquely identify this subfield as indicated in Microsoft Academic (Section 3.2), and the abbreviations are used throughout this paper for brevity. To elaborate, CS, when referred to *as a subfield*, concerns areas such as theoretical computer science (e.g. formal methods). ACM is the world's largest educational and scientific computing society and delivers resources that advance computing as a science and a profession, including to each of these communities.

However, we note that an important area of emerging research in computing – Information Ethics (also known as AI Ethics and equivalent) – is not covered in our survey; nor is the long-established area of Software Engineering. The latter is, for our purposes, classed as a subfield of engineering. And although there have over many decades been nods to ethics, the field of ethics in computer science is still emerging as a truly interdisciplinary field and as a rigorous discipline that, for example, involves experts in philosophical ethics .

### 3.2 Data Collection

Based on the literature surveyed e.g. [25], we have decided to use Microsoft Academic [51] as our data source, as it comes with a programmer-friendly API[3] for programmatic data downloads of citation information. More importantly, Microsoft Academic also provides author and paper metadata, such as research organization and paper category (*topic*), which lets us classify each paper based on their subfield of computer science (Section 3.1).

For each of the 9 subfields in Section 3.1, we query the Microsoft Academic API v1.0 [51] for 20,000 citation records, provided on a best-effort basis using the default parameters of the API. The total of N=20,000 is chosen as it balances the need for a large sample size with due consideration for Microsoft's server resources. Experimentally, values higher than 20,000 results in server time-outs which indicate a high server load; we avoid this to conserve server resources. To further reduce the strain on the server in consideration of the data provider and other fellow users, successful data fetches are limited to no more than two per hour, and the metadata items requested are limited to only a subset of the full metadata available.

The results provided from the Microsoft Academic API are in JSON format, which is then processed in Python for subsequent steps. We firstly perform data deduplication by removing any duplicated citation within- and across-categories, such that any unique citation appears only once within the entire dataset. (Ties are broken in alphabetical order, e.g. a paper which has been dual-classified in the topic 'Artificial Intelligence' and 'Machine Learning' will be included in the former, but not the latter). A grand total of 150,651 citations are obtained after the deduplication process.

### 3.3 Gender Inference: Technology, Caveats, Considerations

By considering the options in our aforementioned literature review on name-to-gender algorithms [45], we have decided to use a two-step process in the interest of reducing costs (in the case of paid online services), while maintaining some degree of transparency to the process (by prioritizing offline methods using published datasets).

Based on the analysis given in [45] as well as initial experimentation with popular Python gender-detection libraries[4], we have chosen *gender-guesser* for the offline option; and keeping in line with extant research methods [25,50], we chose the *genderize.io* web service as the paid online option.

To recap, Section 1 covered broad methodological caveats that we would like to address upfront.

The following algorithm was used in determining gender distributions in each of the particular CS subfields.

1. For each citation, obtain first names of all authors.

2. Authors whose first name consists of a sole initial are discarded as the gender detection algorithm[5] will not work.

3. Each author's first name is first processed with the offline *gender-guesser* [42] Python library, which returns a classification and a classification confidence: {'male', 'mostly male'} (i.e. predicted as male with a higher and a lower degree of confidence respectively), {'female', 'mostly female'} (as before), 'androgynous' (gender neutral), or 'unknown'.

   The difference between the last two is that androgynous names could statistically be in either the male and female name classes, e.g. *Pauley* [42]; whereas unknown names are not found at all in the dataset used by the library.

---

[3] We initially considered the use of the ArXiV repository as it is a popular site hosting preprints for computer science papers. Unfortunately, author first names are provided only as initials (e.g. 'J. Doe' instead of 'Jane Doe'), which render the name-to-gender algorithms ineffective.

[4] Some libraries including gender-detector were promising candidates due to their usage of open datasets, but ultimately were not suitable for our purposes due to performance issues and compatibility issues.
[5] See footnote #3.

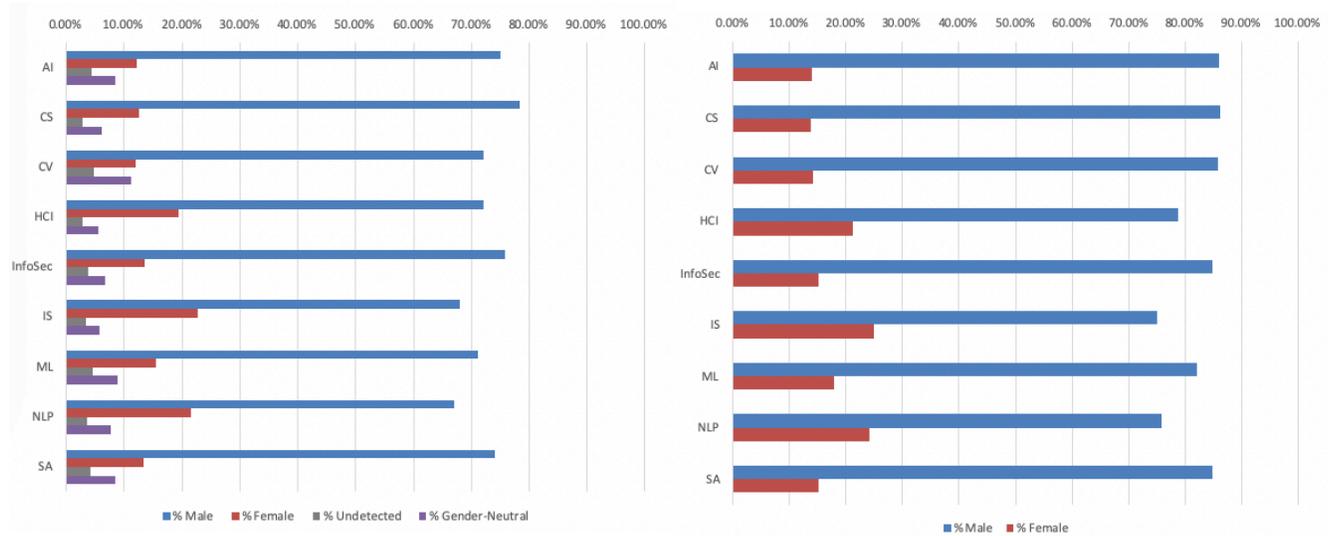

**Figure 1. (Left) Gender distribution of authors by subfield, including gender-neutral names and undetected names; and (Right) Direct comparison between male and female author ratios, excluding gender-neutral and undetected names.**

4. Any names marked 'unknown' by *gender-guesser* are likely to be from a non-anglophone background: hence, based on the track record of *genderize.io* for processing culturally-diverse names, it is used for second-round processing. *Genderize.io* returns either 'male', 'female', or 'unknown'.

5. The final classification from Step 3 (or Step 4 if Step 3 was inconclusive) was then was then used an output point, in the overall aggregation. Repeat Step 1 for all authors per citation, and for all citations in dataset.

Note that we do not remove duplicated names across papers – e.g. if hypothetical author *Shanti Kumar* was present across three papers, we count three separate instances. This presupposes that removing duplications may in fact present a more negative outcome than the current analysis (containing a margin of error) already present; further, we do not wish to dilute the contribution of a single author but instead would want to consider *overall impact*.

## 4 Experimental Results

### 4.1 Statistics on Gender in Subfield

Figure 1 illustrates the overall distribution of the genders resulting from our inference method (Section 3.3): the left inset includes names where gender cannot be inferred reliably by automated techniques (undetected), as well as when a name is deemed to be gender-neutral; and the right inset has gender-neutral terms and undetected names removed, for a direct comparison. The total number of author names range from 42,991 to 65,979 (*mean = 55924.22, s.d. = 7737.88*) per subfield.

The proportion of males outnumber those of females for each of the nine subfields in our study. For the direct comparison case, a simple Analysis of Variance (ANOVA in Microsoft Excel's Data Analysis Toolpak) with *alpha* = 0.05 confirms that the count of males versus females is statistically significant. *(F value = 214.63) > (F critical = 4.49), with p = 1.08774E-10.*

### 4.2 Gender Diversity by Subfield

#### 4.3.1 Less Gender Diverse: AI, CS, InfoSec, CV, ML, SA

From Figure 1(b), we observe that gender representation in publications within a particular subfield is roughly divided into two categories, based on ratio of gender, which we term 'less gender diverse' and 'slight improvement'.

The former is the subject of analysis in this subsection. Accounting for a direct male-to-female comparison, these subfields of AI, CS, InfoSec, CV, ML, and SA have between 10% to 20% female authorship. This translates to an approximate 5:1 ratio of males to females. To make sense of these statistics, we turn to extant literature discussing the state of these subfields, to compare our findings against the actual population of female academics.

AI and ML has been traditionally low in female representation, with "significant differences in machine learning and computer ethics between the United States and the United Kingdom as well as differences in the research focus of papers with female co-authors" [54]. One hypothesis is that subfields with high emphasis on mathematical and scientific techniques (such as ML, CV) suffer more from historical biases, leading to an overrepresentation of males (see Section 5.2). When we examine InfoSec, industry trends in 2016 based on a survey by the *ISC2* cybersecurity professional organisation[8] are such that "…women in the information security profession represent 10% of the global workforce, a percentage that remains unchanged from the 2013 study [but]… 26% of IT professionals worldwide are women" [11].

### 4.3.2 Slight Improvement: NLP, IS, HCI

The second category encompasses the subfields of NLP, IS, and HCI, and has the proportion of female authors ranging from 20% to 30%. These have a nett effect of a roughly 3:1 male:female ratio.

Extant findings from these subfields explain the slight improvement in the male:female ratios. HCI has had issues regarding representation of women which are canvassed by McKay and Buchanan [32]. Via an analysis of OzCHI, the Australian HCI conference venue, these authors claim that "…female representation is quite good, but we need to be cautious to preserve it". In the subfield of NLP, we hypothesise that female representation is higher than the broader umbrella of AI, based on the anecdotal evidence from research addressing *language* bias [55]. To quote Leavy, "…[l]eading thinkers in the emerging field addressing bias in artificial intelligence [specifically language models]… are also primarily female, suggesting that those who are potentially affected by bias are more likely to see, understand and attempt to resolve it" [29].

As for IS, academics in this subfield are in a position to "contribute to addressing the challenge of gender imbalance in the IT profession" [18], noting that issues of gender discrimination in IS have been found as early as 1996, if not earlier [60]. A more detailed discussion on female perceptions of IS and subfields which closely relate to the genesis of computing can be found in Section 5.2.

### 4.3 Algorithmic Limitations

Our study has limitations. The automated nature of gender recognition has technical and methodological caveats as detailed in Section 3.3.. From initial experiments, the margin of 'undetected' names is unacceptably high (ranging from 19.40% to 31.07%) in the absence of using the *Genderize.io* as a second step. This illustrates the fact that existing methods still have a way to go in future research. Human judgement remains preferable over machine analyses and could be performed in future research with smaller samples, or in a hybrid human coding technique assisted by a rule-based system.

## 5 Discussion

### 5.1 Lessons from the Ethics of Care

Our results raise important ethical and social issues. To explore them, we shall use the normative theory known as "ethics of care" or "care ethics" (CE) [22,53]. We choose this ethical approach not because it is the only useful normative theory for exploring such issues, nor because it is beyond criticism (all normative theories are controversial), but because its strong connection to key themes arising from issues pertinent to our study makes it particularly illuminating. Although CE has been applied in some detail to areas like medicine [57] and business [21], it has been less frequently applied to computer science [21,50].

CE arose in the 1970s and 1980s in the context of feminist critiques of male-dominated historical and prevailing ways of doing philosophical ethics. Although CE and feminist ethics [26,58] are distinct—such that, for example, one can be a feminist ethicist without subscribing to CE—CE nevertheless draws heavily on feminist modes of thought that question 'masculine' moral approaches and assumptions that historically were never or were only rarely questioned. CE criticised the then dominant conception of the moral agent as independent, unattached, self-sufficient, unemotional, and rationalistic.

Psychologist Carol Gilligan's seminal early 1980s work *In A Different Voice* brought to light the dominance in ethics discourse of moral values related to this conception of moral agents [17]. These associated values included a preference for relatively unvarying principles and rules, impartiality and detachment, liberal ideals of justice, and contractarian thinking. At around the same time, the philosopher Nel Noddings [39] emphasised the importance to moral agency and experience of human interdependence and of caring and being cared for. The emerging, more relational conception of moral agency and moral life stressed values such as responsiveness, compassion, contextual understanding, and co-dependent relationships. A greater attentiveness to lived experience—as opposed to a detached manner of observation that obscures the details of individuals' lives—was brought to the fore.

Since Gilligan and Noddings, a range of female (and sometimes male, [53]) philosophers have added to the corpus of work on CE, deepening its basis and applying its ideas to a range of contexts and practices. Philosopher Raja Halwani summarises the essence of CE by identifying four of its 'desiderata', namely:

> …the concern with people embedded in contextual relations; attention to areas of life, neglected by some traditional moral philosophy, such as friendship and the family…; the emphasis

---
[8] The organisation conducting this survey is stylised (ISC)².

on the emotive component in ethical engagement; and partiality [20].

Thus, CE foregrounds nurturing relationships, biological and emotional needs, affective engagement or what exponents call "engrossment" [38], and the role of associated contextual features in shaping morality and giving rise to responsibility and obligation. For many care ethicists, these elements spring especially from the lives and historical experiences of women; such so-called "feminine" values and experiences tend to stand in contrast to the experiences of men and the values championed by predominantly male philosophers. In these ways, CE represents a distinct approach to ethics alongside the more established normative theories of deontology, utilitarianism, contract theory, and (to an arguably lesser extent) virtue ethics. However, CE has been criticised on a number of fronts. Broaching several such criticisms, and briefly discussing how CE can respond to them, will help us to apply this approach to the findings of this study.

CE might be criticised for essentializing gender, overlooking gender diversity and fluidity, and stereotyping men and women. For example, some men embody 'feminine' values like nurturing and compassion just as some women embody 'masculine' values like moral rationalism and extreme impartiality. However, although CE grew from and encompasses feminist critiques of male-centred ethics, it may reject essentialism, accept non-binary views of gender, and agree that 'feminine' values of caring are not the exclusive preserve of women. Indeed, it may champion these positions while stressing that women nonetheless are often well-placed to identify and respond to the values at the heart of the CE approach. In Noddings' words, there are "centuries of experience more typical of women than men" [39]. Such experience, of course, includes not only the experience of the mother-child relation, but the experience of caring more generally [44].

The patriarchal conditions of women's historical caring (including motherhood) may raise the concern that the caring outlook is not always laudable but rather may be a function of oppression and an associated distortion of perspective. Entrenched power imbalances, it might be suggested, could have led to forms of moral blindness centring on a problematic valourising of caring relations [8]. Yet a CE proponent may reply that these historical and, moreover, persisting inequalities in power may, on the contrary, often give women—and, we should also stress, women at various intersections, such as women of colour and women with disabilities—greater insight into a range of moral matters, including the unfairness of many circumstances and caring relations occupied predominantly by women, the oppression of people of different sexualities and genders, and the needs of individuals and groups who are marginalised and especially vulnerable. This is not to say that men cannot also adopt such perspectives and the relevant forms of moral attentiveness; it is rather to note that ethical insight into certain states of affairs can sometimes be sharpened by cultural, historical, and biological circumstances. At the same time, we can say that from the point of view of care ethicists, men too have reason to adopt a CE position.

Given these points, it is perhaps somewhat ironic that CE has been criticised for lacking the resources to give guidance on political and moral questions regarding those to whom we do not have special and partial relations. Nonetheless, it is a serious question. Can CE, then, say anything significant about *justice* outside of those paradigmatic relations of care? [13,59] There are many individuals more distant from us both geographically and personally who are nevertheless particularly vulnerable. CE, its proponents may say, can both recognise and highlight the situations of such people and the moral necessity for them to receive care just as those personally close to us need and deserve care. Furthermore, the ethical attentiveness and responsiveness to need that is rightly cultivated and honoured in more personal caring contexts can in some form be extended to strangers, be they disenfranchised fellow citizens or people from foreign places. The caring attitude can also be extended to, say, academic colleagues who have needs and require support and even sometimes nurturing. Therefore, CE need not displace or overlook the promotion of justice for more 'distant' others, but rather can (arguably) recognise, inform, and deepen the notion of justice. Again, people of any sex or gender can adopt this CE viewpoint.

## 5.2 Ethics of Care and Computer Science

Having briefly outlined some of the features, problems, and strengths of CE, we can now apply it to our findings. Our study suggests that women are strongly underrepresented in all subfields of computer sciences, in some even more than in others. From a CE perspective, this is problematic for several reasons which we will now briefly discuss. In the first place, there is the issue of justice or fairness for women themselves. Many women have historically had strong interests in pursuing careers across the various divisions of computer science. However, due to historical and persisting power imbalances which may include both overt discrimination and subtle and implicit biases, it is often harder for women than it is for men to enter computer sciences and, once there, to climb institutional ladders. Interestingly, it was not always this way. The first analytical machine, the precursor to the computer, was created by Ada Lovelace [48]. Computing was predominantly populated by women until it became a 'profession' and was paid a proper salary - around the mid-1980's[9] [23].

In essence, critical factors dissuading participation of females in computer science in the beginning of the 'pipeline' include the change in perception from the 1970s, from a field perceived to be "… more clerical in nature" to its redefinition as "a science… [which] distanced itself from skill sets traditionally thought to be well suited to women and sought to align itself with other science fields like engineering that had strong masculine connotations" [46] (citing [14]). Existing biases against women in STEM-based academic jobs may draw them towards leaving the 'academic pipeline' [2] (see also Section 2). This effect is partially offset by countervailing factors such as the promotion of computing skills in mathematics [56] or affirmative action policies to close the gender gap [7].

---

[9] A particularly relevant quote by Steve Henn explains this: *"A lot of computing pioneers — the people who programmed the first digital computers — were women. And for decades, the number of women studying computer science was growing faster than the number of men. But in 1984, something changed"* [23]. See also https://www.history.com/news/coding-used-to-be-a-womans-job-so-it-was-paid-less-and-undervalued

Nonetheless, as our results show, it appears to be the case that women are particularly excluded from certain areas of computer sciences like CV, AI, and InfoSec. Moral attentiveness and responsiveness to women's important needs are qualities and behaviours that a CE viewpoint can endorse and promote. In addition, CE can, as we explained, recognise the importance of forms of justice that regard others as deserving of care, and, in this specific context, as individuals whose talents and aspirations should be appreciated and nurtured. As we noted, CE is strongly oriented to recognising and responding to historical and persisting disadvantage and vulnerability.

Furthermore, CE suggests that, again due to historical and persisting circumstances, women, including those with lived experience at various social and political 'intersections',often have particularly keen moral insights into matters related to care, gender, vulnerability, marginalization, and so on. Again, making this claim is not necessarily to fall prey to gender essentialism or the stereotyping of any particular gender; it is rather to register the potential effects of context and circumstance on the moral attentiveness and responsiveness of certain situated individuals. Insofar as (some) women bring these qualities and perspectives into areas of computer science, they may help to expand the moral awareness of their colleagues, including those who are more established and in positions of power. And this may be of benefit to future women and other marginalised groups who want to work in those fields.

Increasing the chances of the entry of other moral perspectives and experiences into computer sciences is likely to generate benefits to others beyond those who wish to work in computer sciences. This could include both the subjects of computer sciences research and wider groups of individuals, including disenfranchised and marginalised people. Moral qualities prized by CE are often important in regard to the rightful treatment of individuals in research and experiments, especially those who are more vulnerable, such as transgender individuals, children, and some people with disabilities. HCI researchers may, to take just one example, study the needs of older people, including those with dementia, for attaining degrees of digital literacy and enrichment [62]. Due in some part to implicit biases, older adults are an example of a group that tends to get overlooked in relation to new technologies.

Care ethicists themselves have often focused on various groups who have been socially marginalised, subjected to prejudice, and, furthermore, relatively neglected by mainstream normative theories. Eva Kittay, for example, has used CE to stress the moral necessity of caring in the right way not just for individuals who are capable of achieving a full range of human flourishing, but also and vitally those with severe cognitive impairments whose opportunities in life are comparatively, and sometimes profoundly, limited [27]. Kittay highlights the value of such caring relations to both the carer and the cared-for. For thinkers like her, CE asks us to be just as attentive to the needs of people struck by misfortune as to the interests of people who have higher degrees of self-sufficiency, autonomy, and independence [41].

Both CE and the moral experiences and perspectives of many women lend themselves to a caring-style concern and sense of justice for vulnerable individuals in the community who may be more broadly affected by the computer sciences. Sometimes this will involve a sensitivity to the effects of computer sciences on women themselves. For example, close attention must be paid to technologies which require the training of algorithms on data sets that could introduce gender biases with potential negative consequences for women in the general public. The same also goes for technologies that could harm various marginalised and vulnerable groups of people (and, as for example ecofeminists recognize, nonhuman animals).

Identification and active correction of a range of ethico-social problems is not only important because of the harm done to various individuals. Ameliorating existing inequality also has the potential to improve the quality of the science itself. Just as being attentive to the song of female as well as male birds — a dimension of animal behaviour overlooked because of the bias of male researchers — enriches and improves ornithology, so too does heightened attentiveness to biases, faulty assumptions, and prejudice potentially enrich, broaden, and improve computer sciences. This applies not only to the more human-centred subfields like HCI, but also to more technical (and, as our results show, especially male-dominated) subfields like CV and AI. Therefore, creating the conditions for greater gender equality (and other kinds of equality) across the subfields of computer sciences should be seen less as a threat than as an opportunity for enhancing the rigour and value of the discipline. On top of the ethical arguments, this last point about the quality of the disciplines provides further support and additional leverage for making changes to the current system to advance gender equity.

### 5.3 Recommendations and Professional Responsibilities

The CE framework supports not only the passive avoidance of, say, discriminatory hiring practices, but also the active nurturing of individuals and the modification of institutional attitudes and structures that prevent women (and other minoritised groups) from fairly occupying roles in the various subfields of computer science. Furthermore, our discussion of how CE applies to computer science can be used to justify and reinforce existing professional standards.

The Association for Computing Machinery (ACM) Code of Ethics and Professional Conduct [4], for example, contains several ethical principles relevant to our discussion of gender (and other kinds of) diversity. Thus, Principle 1.4 of the ACM Code says that

> Computing professionals should foster fair participation of all people, including those of underrepresented groups. Prejudicial discrimination on the basis of age, color, disability, ethnicity, family status, gender identity, labor union membership, military status, nationality, race, religion or belief, sex, sexual orientation, or any other inappropriate factor is an explicit violation of the Code. The use of information and technology may cause new, or enhance existing, inequities. Technologies and practices should be as inclusive and accessible as possible and computing professionals should take action to avoid creating systems or technologies that disenfranchise or oppress people. Failure to design for inclusiveness and accessibility may constitute unfair discrimination. [4]

Indeed, the ACM Code even says (Principle 1.1) that when "the interests of multiple groups conflict, the needs of those less advantaged should be given increased attention and priority." [4] The Code then, calls for resolute action to be taken on behalf of individuals facing injustice and discrimination. Such a strong stance is supported by a CE approach, with its particular emphasis upon people in need of care and justice. Furthermore, the ACM calls upon its members to (Principle 2.1) "strive to achieve high quality in both the processes and products of professional work" [4]. This provides another reason for striving to remove disadvantage and injustice since, as we have argued, underrepresentation of women (and other groups) can have a negative effect on the quality and breadth of the work done in the subfields of computer science.

Given the standards to which the ACM aspires under its *Code of Ethics and Professional Conduct*, and given the arguments we have presented from the ethics of care, we would argue that the computing community has responsibilities and caring duties to promote and support those standards. One concrete way to promote this is for the ACM to monitor compliance with its principles and to set up a dashboard of compliance against which computer science communities can measure themselves annually, providing accountability within their own community and to the computer science community at large. This includes metrics of performance and inclusion not only of gender, but also of race, disability, class, sexuality and numerous other minoritised groups. Providing such metrics will raise awareness and comparison between communities, hopefully also leading to sharing of best practice and changes in behaviour. Meanwhile, those communities already working on diversity and inclusion and contemplating governance could share their experience and strategies with other communities. In addition, a best practice guide might be facilitated by the ACM. These are just a sample of the concrete steps the ACM should consider. Which other detailed responses should be pursued depends on further research, including hearing the views of a diverse range of stakeholders. Our key point in this preliminary study, however, is that such concrete steps are urgent and necessary, and that they should be guided by the sorts of attentiveness to and engagement with minoritized individuals that CE so clearly brings out.

## 6 Future Work and Conclusion

This study is the first step in a bigger project to identify outputs and cultures relating to fairness, accountability, and transparency within the different computer science communities. To move this project forward, it would be helpful to investigate what systems, research, and active members address these issues, and to test the hypothesis that more diverse communities will, in fact, have more advanced systems, academic work, and conversations about fairness, accountability and transparency. If this hypothesis is not correct, it would be useful to explore what helps to advance these conversations and considerations within individual communities. Our hypothesis is based on research in other fields that have indicated that diversity is key to good research.

Using this study as a steppingstone, we intend to undertake further qualitative research to analyse what works in the various communities, what does not, and how these successes and failures can be better shared. The next step involves qualitative research within each of the communities to ask minoritised members about their lived experience and the cultures within each field, including successes and failures, and what we can learn from both to achieve the standards set out by the ACM code of ethics. Finally, we would invite other scholars to investigate how we could fulfil our ethical duties of care to the computer science and broader community.

## REFERENCES

1. Alison Adam and Jacqueline Ofori-Amanfo. 2000. Does gender matter in computer ethics? *Ethics and information technology* 2, 1: 37–47.
2. R. Agrawal, P. Goodwill, N. Judge, M. Sego, and A. Williams. 2001. *The shortage of female computer science faculty at Stanford University*. Stanford University. Retrieved from https://cs.stanford.edu/people/eroberts/courses/cs181/projects/2000-01/women-faculty/intro.html
3. P. M. Asaro. 2019. AI Ethics in Predictive Policing: From Models of Threat to an Ethics of Care. *IEEE Technology and Society Magazine* 38, 2: 40–53.
4. Association for Computing Machinery. 2016. Code of Ethics. *ACM Ethics*. Retrieved October 8, 2020 from https://ethics.acm.org/code-of-ethics/
5. Abeba Birhane and Olivia Guest. 2020. Towards decolonising computational sciences. *arXiv [cs.CY]*. Retrieved from http://arxiv.org/abs/2009.14258
6. Anne Boring. 2017. Gender biases in student evaluations of teaching. *Journal of public economics* 145: 27–41.
7. Carmen Botella, Silvia Rueda, Emilia López-Iñesta, and Paula Marzal. 2019. Gender Diversity in STEM Disciplines: A Multiple Factor Problem. *Entropy* 21, 1: 30.
8. Claudia Card. 1996. *The Unnatural Lottery: Character and Moral Luck*. Temple University Press.
9. Marc Cheong. 2013. Inferring Social Behavior and Interaction on Twitter by Combining Metadata about Users & Messages. Clayton School of Information Technology, Monash University.
10. Sapna Cheryan, Victoria C. Plaut, Paul G. Davies, and Claude M. Steele. 2009. Ambient belonging: how stereotypical cues impact gender participation in computer science. *Journal of personality and social psychology* 97, 6: 1045–1060.

11. Eleanor Dallaway. 2016. Closing the gender gap in cybersecurity. *CREST*. Retrieved October 8, 2020 from https://www.crest-approved.org/wp-content/uploads/CREST-Closing-the-Gender-Gap-in-Cyber-Security.pdf
12. Catherine D'Ignazio and Lauren Klein. 2020. 1. The Power Chapter. In *Data Feminism*.
13. Daniel Engster. 2009. *The heart of justice*. Oxford University Press, London, England.
14. Nathan L. Ensmenger. 2012. *The Computer Boys Take Over: Computers, Programmers, and the Politics of Technical Expertise*. MIT Press.
15. Virginia Eubanks. 2018. *Automating Inequality: How High-Tech Tools Profile, Police, and Punish the Poor*. St. Martin's Publishing Group.
16. Diana Franklin. 2013. A practical guide to gender diversity for computer science faculty. *Synthesis lectures on professionalism and career advancement for scientists and engineers* 1, 2: 1–81.
17. Carol Gilligan. 1993. *In a Different Voice: Psychological Theory and Women's Development*. Harvard University Press.
18. Elena Gorbacheva, Jenine Beekhuyzen, Jan vom Brocke, and Jörg Becker. 2019. Directions for research on gender imbalance in the IT profession. *European journal of information systems: an official journal of the Operational Research Society* 28, 1: 43–67.
19. Casey D. Haines, Evangeline M. Rose, Karan J. Odom, and Kevin E. Omland. 2020. The role of diversity in science: a case study of women advancing female birdsong research. *Animal behaviour* 168: 19–24.
20. Raja Halwani. 2003. Care Ethics and Virtue Ethics. *Hypatia* 18, 3: 161–192.
21. Maurice Hamington and Maureen Sander-Staudt. 2011. *Applying Care Ethics to Business*. Springer Science & Business Media.
22. Virginia Held. 2006. *The Ethics of Care*. Oxford University Press.
23. Steve Henn. 2014. When women stopped coding. *Planet Money: NPR*. Retrieved October 8, 2020 from https://www.npr.org/sections/money/2014/10/21/357629765/when-women-stopped-coding
24. Luke Holman, Devi Stuart-Fox, and Cindy E. Hauser. 2018. The gender gap in science: How long until women are equally represented? *PLoS biology* 16, 4: e2004956.
25. Junming Huang, Alexander J. Gates, Roberta Sinatra, and Albert-László Barabási. 2020. Historical comparison of gender inequality in scientific careers across countries and disciplines. *Proceedings of the National Academy of Sciences of the United States of America* 117, 9: 4609–4616.
26. Alison M. Jaggar. 1991. *Feminist ethics: Projects, problems, prospects*. University of Kansas.
27. Eva Feder Kittay. 1999. *Love's labor: Essays on women, equality and dependency*. Routledge.
28. Vivian Anette Lagesen. 2008. A Cyberfeminist Utopia?: Perceptions of Gender and Computer Science among Malaysian Women Computer Science Students and Faculty. *Science, technology & human values* 33, 1: 5–27.
29. S. Leavy. 2018. Gender bias in artificial intelligence: The need for diversity and gender theory in machine learning. In *2018 IEEE/ACM 1st International Workshop on Gender Equality in Software Engineering (GE)*, 14–16.
30. Lillian MacNell, Adam Driscoll, and Andrea N. Hunt. 2015. What's in a name: Exposing gender bias in student ratings of teaching. *Innovative Higher Education* 40, 4: 291–303.
31. Lisa L. Martin. 2016. Gender, Teaching Evaluations, and Professional Success in Political Science. *PS, political science & politics* 49, 2: 313–319.
32. Dana McKay and George Buchanan. 2019. Shaking the tree. In *Proceedings of the 31st Australian Conference on Human-Computer-Interaction*. https://doi.org/10.1145/3369457.3369504
33. Stefania Milan and Emiliano Treré. 2019. Big Data from the South(s): Beyond Data Universalism. *Television & New Media* 20, 4: 319–335.
34. Kristina M. W. Mitchell and Jonathan Martin. 2018. Gender Bias in Student Evaluations. *PS, political science & politics* 51, 3: 648–652.
35. Blanca Myers. 2018. Women and Minorities in Tech, By the Numbers. *Wired*. Retrieved September 29, 2020 from https://www.wired.com/story/computer-science-graduates-diversity/
36. J. Nathan Matias. 2014. How to Ethically and Responsibly Identify Gender in Large Datasets. *Mediashift*. Retrieved October 5, 2020 from http://mediashift.org/2014/11/how-to-ethically-and-responsibly-identify-gender-in-large-datasets/
37. Saif M Mohammad. 2020. Gender Gap in Natural Language Processing Research: Disparities in Authorship and Citations. In Proceedings of the 58th Annual Meeting of the Association for Computational Linguistics. Association for Computational Linguistics, 7860–7870. https://www.aclweb.org/anthology/2020.acl-main.702v2.pdf
38. National Science Foundation. Nsf.Gov - NCSES science and engineering degrees: 1966–2010. *US National Science Foundation (NSF)*. Retrieved October 7, 2020 from https://www.nsf.gov/statistics/nsf13327/content.cfm?pub_id=4266&id=2
39. Nel Noddings. 2013. *Caring: A Relational Approach to Ethics and Moral Education*. Univ of California Press.
40. Kathryn Norlock. 2019. Feminist Ethics. *The Stanford Encyclopedia of Philosophy*. Retrieved from https://plato.stanford.edu/archives/sum2019/entries/feminism-ethics/
41. Martha C. Nussbaum. 2006. *Frontiers of justice*. Belknap Press, London, England.
42. Israel Saeta Pérez. 2020. *gender-guesser*. Retrieved October 5, 2020 from https://pypi.org/project/gender-guesser/
43. Andrew S. Rosen. 2018. Correlations, trends and potential biases among publicly accessible web-based student evaluations of teaching: a large-scale study of RateMyProfessors.com data. *Assessment & Evaluation in Higher Education* 43, 1: 31–44.
44. Sara Ruddick. 1989. Maternal Thinking: Toward a Politics of Peace. *New York: Ballantine*.
45. Lucía Santamaría and Helena Mihaljević. 2018. Comparison and benchmark of name-to-gender inference services. *PeerJ Computer Science* 4: e156.
46. Linda J. Sax, Kathleen J. Lehman, Jerry A. Jacobs, M. Allison Kanny, Gloria Lim, Laura Monje-Paulson, and Hilary B. Zimmerman. 2017. Anatomy of an Enduring Gender Gap: The Evolution of Women's Participation in Computer Science. *The Journal of higher education* 88, 2: 258–293.
47. Linda J. Sax, Hilary B. Zimmerman, Jennifer M. Blaney, Brit Toven-Lindsey, and Kathleen J. Lehman. 2017. Diversifying undergraduate computer science: The role of department chairs in promoting gender and racial diversity. *Journal of women and minorities in science and engineering* 23, 2: 101–119.
48. Miranda Seymour. 2018. *In Byron's Wake*. Simon & Schuster, London, England.
49. Allison K. Shaw and Daniel E. Stanton. 2012. Leaks in the pipeline: separating demographic inertia from ongoing gender differences in academia. *Proceedings. Biological sciences / The Royal Society* 279, 1743: 3736–3741.
50. Alicia Shen and Ione Fine. 2018. Perish not publish? New study quantifies the lack of female authors in scientific journals. *The Conversation*. Retrieved October 5, 2020 from http://theconversation.com/perish-not-publish-new-study-quantifies-the-lack-of-female-authors-in-scientific-journals-92999
51. Arnab Sinha, Zhihong Shen, Yang Song, Hao Ma, Darrin Eide, Bo-June (paul) Hsu, and Kuansan Wang. 2015. An Overview of Microsoft Academic Service (MAS) and Applications. In *Proceedings of the 24th International Conference on World Wide Web* (WWW '15 Companion), 243–246.
52. Rebecca Slayton. 2013. *Arguments that Count: Physics, Computing, and Missile Defense, 1949-2012*. MIT Press.
53. Michael Slote. 2007. *The Ethics of Care and Empathy*. Routledge.
54. Konstantinos Stathoulopoulos and Juan C. Mateos-Garcia. 2019. Gender diversity in AI research. *SSRN Electronic Journal*. https://doi.org/10.2139/ssrn.3428240
55. Tony Sun, Andrew Gaut, Shirlyn Tang, Yuxin Huang, Mai ElSherief, Jieyu Zhao, Diba Mirza, Elizabeth Belding, Kai-Wei Chang, and William Yang Wang. 2019. Mitigating gender bias in Natural Language Processing: Literature review. *arXiv [cs.CL]*. Retrieved from http://arxiv.org/abs/1906.08976
56. Mike Thelwall, Carol Bailey, Meiko Makita, Pardeep Sud, and Devika P. Madalli. 2019. Gender and research publishing in India: Uniformly high inequality? *Journal of informetrics* 13, 1: 118–131.
57. R. Tong. 1998. The ethics of care: a feminist virtue ethics of care for healthcare practitioners. *The journal of medicine and philosophy* 23, 2: 131–152.
58. Rosemarie Tong. 1995. *Feminine and Feminist Ethics*. Wadsworth Publishing Company.
59. Joan C. Tronto. 2020. *Moral boundaries*. Routledge.
60. Gregory E. Truman and Jack J. Baroudi. 1994. Gender differences in the information systems managerial ranks: An assessment of potential discriminatory practices. *MIS quarterly: management information systems* 18, 2: 129.
61. Lucy Lu Wang, Gabriel Stanovsky, Luca Weihs, and Oren Etzioni. 2019. Gender trends in computer science authorship. *arXiv [cs.DL]*. Retrieved from http://arxiv.org/abs/1906.07883
62. Jenny Waycott, Amee Morgans, Sonja Pedell, Elizabeth Ozanne, Frank Vetere, Lars Kulik, and Hilary Davis. 2015. Ethics in evaluating a sociotechnical intervention with socially isolated older adults. *Qualitative health research* 25, 11: 1518–1528.